\documentclass[11pt, superscriptaddress, groupedaddress, aps, pra, floatfix]{revtex4-2}

\usepackage{graphicx}
\usepackage{color}
\usepackage{amsmath}
\usepackage{amsfonts}
\usepackage{amssymb}
\usepackage{bbold}
\usepackage{url}
\usepackage{mathtools}

\begin{document}

\title{Phase-sensitive representation of Majorana stabilizer states}

\author{Tomislav Begu\v{s}i\'{c}}
\altaffiliation{Present address: Institute of Physical and Theoretical Chemistry, University of W\"urzburg, 97074 W\"urzburg, Germany}
\author{Garnet Kin-Lic Chan}
\email{gkc1000@gmail.com}
\affiliation{Division of Chemistry and Chemical Engineering, California Institute of Technology, Pasadena, California 91125, USA}

\date{\today}

\begin{abstract}
Stabilizer states hold a special place in quantum information science due to their connection with quantum error correction and quantum circuit simulation. In the context of classical simulations of many-body physics, they are an example of states that can be both highly entangled and efficiently represented and transformed under Clifford operators. Recently, Clifford operators have been discussed in the context of fermionic quantum computation through their extension, the Majorana Clifford group. Here, we document the phase-sensitive form of the corresponding Majorana stabilizer states, as well as the algorithms for computing their amplitudes, their inner products, and update rules for transforming Majorana stabilizer states under Majorana Clifford gates.
\end{abstract}

\maketitle

\section{Introduction}

The classical simulation of quantum systems depends strongly on the representation of quantum states. In the context of qubit-based systems, an interesting example of states that are both highly entangled and efficiently representable are stabilizer states \cite{Gottesman:1997, Gottesman:1998}, which are formed by applying Clifford gates to an initial computational basis state. Importantly, they can be difficult to represent using other common strategies, such as computational basis expansion or tensor networks.

In fermionic systems, and especially in quantum chemistry, the connection with concepts from quantum information science has typically proceeded by transforming into the qubit picture \cite{Mishmash_Mezzacapo:2023, Schleich_AspuruGuzik:2023, Projansky_Whitfield:2025}. This can lead to a loss of interpretability, given that typical transformations, such as the Jordan-Wigner transformation, map local fermionic operators into non-local Pauli operators. Yet, discussions of stabilizer states and Clifford operators in fermionic systems are increasingly common, with applications ranging from quantifying non-stabilizerness of various fermionic states \cite{collura2025nonstabilizernessfermionicgaussianstates, Bera_Schiro:2025, Sierant_Turkeshi:2026} to identifying Clifford transformations that simplify the quantum-chemical electronic structure problem \cite{Mishmash_Mezzacapo:2023, Fu_Guo:2025}. Alternatively, Majorana operators can serve as fermionic operators that are unitary and also retain locality of the original creation and annihilation operators. Majorana operators are essential in the discussion of fermionic quantum computation \cite{Bravyi2002, Bravyi_2010, Vijay2015, Plugge_Egger:2016,  OBrien_Akhmerov:2018, Litinski_Oppen:2018, McLauchlan2024}. These operators can be used to define Majorana Clifford gates, a fermionic analogue of qubit Clifford operators, with the property that they transform Majoranas into Majoranas. While the examples of Majorana Clifford operators appeared already in the early work of Bravyi and Kitaev \cite{Bravyi2002} and have been employed since in the context of fermionic quantum computation \cite{Hyart_Beenakker:2013, OBrien_Akhmerov:2018}, they have been explicitly discussed only in recent works under the names of ``logical braid'' operators \cite{McLauchlan2022} and fermionic (or Majorana) Clifford operators \cite{Mudassar_Gottesman:2024,Bettaque_Swingle:2024, mudassar2025faulttolerantoperationsmajoranabased, magoulas2025cliffordtransformationsfermionicquantum}. Majorana Clifford transformations of Majorana operators can be efficiently simulated, which motivates, for example, sparse Majorana representations for the classical simulation of fermionic quantum expectation values \cite{miller2025simulationfermioniccircuitsusing, rudolph2026thermalstatesimulationpauli}. Majorana Clifford operators have also been used to improve Trotter-based quantum simulation of fermionic systems \cite{gandon2026stabilizerbasedquantumsimulationfermion}.

Based on Majorana Clifford gates, one can also define Majorana stabilizer states, which are the topic of this work. In order to use them as a basis for representing arbitrary fermionic quantum states, we need a phase-sensitive representation of Majorana stabilizer states. For context, the analogous, qubit-based stabilizer state can be represented either without phase information \cite{Gottesman:1997, Aaronson_Gottesman:2004}, which is sufficient for evaluating expectation values and sampling from stabilizer states, or including the global phase, which is needed to represent general states as superpositions of stabilizer states. Phase-sensitive representations, such as the CH form \cite{Bravyi_Howard:2019}, the affine form \cite{Dehaene2003, VanDenNest:2010}, and the graph state formalism \cite{Hein2004, Anders2006, Kerzner:2021, Gosset2024}, have been used in methods for simulating quantum circuits dominated by Clifford gates \cite{Bravyi_Gosset:2016, Bravyi_Howard:2019, Pashayan_Bartlett:2022}. Here, we follow the CH form as presented in Ref.~\cite{Bravyi_Howard:2019} to define a phase-sensitive representation of Majorana stabilizer states. We provide algorithms for updating Majorana stabilizer states under Majorana Clifford operators, for computing probability amplitudes, and for computing inner products of two Majorana stabilizer states. These algorithms are sufficient to operate with fermionic states represented as superpositions of Majorana stabilizer states.

\textit{Notation.} Let us introduce the notation that we will use throughout the manuscript. We will reserve $c$ and $\tilde{c}$ for denoting Majorana operators, $a$ ($a^{\dagger}$) for the fermionic annihilation (creation) operators, and we will use a short form $|0\rangle = |000 \dots 0\rangle$ to denote the vacuum state of a $n$-site fermionic system. For a binary vector $x$, we will call a binary vector $\bar{x}$ its parity representation with definition
\begin{equation}
    \bar{x}_k = \sum_{j=0}^{k} x_j,
\end{equation}
and a binary value $\pi(x) = \sum_{j=0}^{n-1} x_j \in \{0, 1\}$ the parity of $x$. All addition of binary numbers is assumed to be mod 2 (logical exclusive OR operation or addition on $\mathbb{Z}_2$), i.e., we will not have a special notation for mod 2 addition. The number of nonzero elements of a binary array $x$ will be denoted by $|x|$. We will also use $e_k$ to denote a binary vector with zeros everywhere except at position $k$, where the value is 1.

\section{Majorana operators}

Single-site Majorana operators are defined through
\begin{equation}
c_k = a_k + a_k^{\dagger}, \quad \tilde{c}_k = i(a_k^{\dagger} - a_k),
\end{equation}
where $a_k^{\dagger}$ and $a_k$ are standard fermionic creation and annihilation operators. In our notation, we deviate from a more conventional approach to label all Majorana operators with $c$ (or $\gamma$), where the indices run from $0$ to $2n-1$ \cite{Bravyi2002, Mudassar_Gottesman:2024}. We chose the above notation to easily keep track of the fermionic site $k$ on which the Majorana operators $c_k$ and $\tilde{c}_k$ act \cite{Surace_Tagliacozzo:2022}. 

Majorana operators satisfy the following anticommutation relations:
\begin{equation}
\{\gamma_j, \gamma^{\prime}_k \} = 2 \delta_{jk} \delta_{\gamma, \gamma'}, \quad \gamma, \gamma' \in \{c, \tilde{c}\}.
\end{equation}
They are also Hermitian and unitary:
\begin{equation}
\gamma^{\dagger}_k = \gamma_k, \quad \gamma_k^2 = I, \quad \gamma \in \{c, \tilde{c}\}
\end{equation}

A general $n$-site Majorana operator (Majorana string) can be defined as
\begin{equation}
\Gamma = i^{\varphi} \prod_{k=0}^{n-1} p_k^{z_k} c_k^{x_k}, \label{eq:MajoranaString}
\end{equation}
where 
\begin{equation}
    p_k = i \tilde{c}_k c_k
\end{equation}
with (anti)commutation relations
\begin{equation}
    [p_j, p_k] = 0, \quad [c_j, p_k] = 0\, (j \neq k), \quad \{c_k, p_k\} = 0,
\end{equation}
$z_k, x_k \in \{0, 1\}$, and $\varphi \in \{0, 1, 2, 3\}$. This representation of Majorana operators can be used to store them in a computer and perform operations, such as checking if two Majorana operators commute or taking their product. Majorana strings are clearly unitary and either Hermitian or anti-Hermitian. We note that an alternative representation of Majorana operators has been used in Ref.~\cite{Mudassar_Gottesman:2024} (see also a slightly different version in \cite{Bettaque_Swingle:2024}),
\begin{equation}
    \Gamma = i^{r} \prod_{k=0}^{n-1} c_k^{m_{2k}} \tilde{c}_k^{m_{2k+1}},
\end{equation}
which stores the operator using a phase $r$ and a binary vector $m$ of length 2$n$. The two representations are related by
\begin{equation}
    \varphi = r + |m| + 2 m_{e} \cdot m_{o}, \quad z = m_{o}, \quad x = m_{e} + m_{o},
\end{equation}
where $m_e = (m_0, m_2, \dots, m_{2n-2})$ is the vector of even elements of $m$ and $m_o = (m_1, m_3, \dots, m_{2n-1})$ is the vector of odd elements of $m$.

The $(\varphi, z, x)$ representation used in this work is closely related to the symplectic representation of Pauli operators. For example, we can verify if an operator is diagonal in the computational basis by simply checking that all $x_k=0$. We will refer to such operators as p-type operators. Similarly, Majorana operators with an even number $|x|$ of nonzero elements in $x$ are of even parity and those with odd $|x|$ are of odd parity. Below, we provide explicit expressions for operations on Majorana strings in the $(\varphi, z, x)$ form.

\subsection{Action on a computational basis state}

Let us assume we are given $\Gamma$ defined by $(\varphi, z, x)$ and a computational basis state $|s\rangle$. Using
\begin{eqnarray}
    c_k |s\rangle &=& (-1)^{\sum_{j=0}^{k-1}s_k} |s + e_k\rangle,\\
    p_k |s\rangle &=& (-1)^{s_k} |s\rangle,
\end{eqnarray}
we have
\begin{eqnarray}
    \Gamma |s\rangle &=& i^{\varphi} \prod_{k=0}^{n-1} p_k^{z_k} c_k^{x_k} |s\rangle \\
    &=& i^{\varphi} (-1)^{s \cdot \bar{x} + \pi(x)\pi(s)} \prod_{k=0}^{n-1} p_k^{z_k}  |s+x\rangle \\
    &=& i^{\varphi} (-1)^{(s+x)\cdot z + s \cdot \bar{x} + \pi(x)\pi(s)} |s+x\rangle,
    \label{eq:gamma_times_s}
\end{eqnarray}
where $e_k$ is a binary vector with zeros everywhere except at position $k$, $\bar{x}$ is the parity representation of $x$, $\pi(x) = \sum_{j=0}^{n-1} x_j$ is the parity of $x$ (see \textit{Notation}).

\subsection{Product of Majorana operators}

Given Majorana operators $\Gamma$ and $\Gamma'$, the product is given by
\begin{eqnarray}
\Gamma \Gamma^{\prime} 
&=&i^{\varphi + \varphi^{\prime}} \prod_{k=0}^{n-1} p_k^{z_{k}} c_k^{x_{k}} \prod_{k=0}^{n-1} p_k^{z_{k}^{\prime}} c_k^{x_{k}^{\prime}} \\
&=&i^{\varphi + \varphi^{\prime}} (-1)^{z' \cdot x} \prod_{k=0}^{n-1} p_k^{z_{k} + z_{k}^{\prime}} c_k^{x_{k}} \prod_{k=0}^{n-1} c_k^{x_{k}^{\prime}} \\
&=&i^{\varphi + \varphi^{\prime}} (-1)^{z' \cdot x + \sum_{k=0}^{n-1} \sum_{j=k+1}^{n-1} x_{k}' x_j }\prod_{k=0}^{n-1} p_k^{z_{k} + z_{k}^{\prime}} c_k^{x_{k} + x_{k}^{\prime}} \\
&=&i^{\varphi + \varphi^{\prime}} (-1)^{z^{\prime}\cdot x + x^{\prime} \cdot \bar{x} + \pi(x^{\prime})\pi(x)} \prod_{k=0}^{n-1} p_k^{z_{k} + z_{k}^{\prime}} c_k^{x_{k} + x_{k}^{\prime}}.
\label{eq:MajoranaProduct}
\end{eqnarray}
Therefore, the parameters of the product $\Gamma \Gamma'$ are $(\varphi + \varphi' + 2(z^{\prime}\cdot x + x^{\prime} \cdot \bar{x} + \pi(x^{\prime})\pi(x)), z+z', x+x')$.

\subsection{Commutation of Majorana operators}

Using the expression derived above, we can conclude that
\begin{equation}
    \Gamma \Gamma' = (-1)^{z^{\prime}\cdot x + x^{\prime} \cdot \bar{x} + z\cdot x^{\prime} + x \cdot \bar{x}^{\prime}} \Gamma'\Gamma,
\end{equation}
i.e., the Majorana operators commute if $z^{\prime}\cdot x + x^{\prime} \cdot \bar{x} + z\cdot x^{\prime} + x \cdot \bar{x}^{\prime} = 0$. We can further avoid evaluating the parity representations of $x$ and $x'$ by rewriting $x^{\prime} \cdot \bar{x} + x \cdot \bar{x}^{\prime} = x\cdot x' + \pi(x) \pi(x')$, which yields
\begin{equation}
    \Gamma\, \text{anticommutes with}\, \Gamma^{\prime} = z^{\prime} \cdot x + x^{\prime} \cdot (z + x) + \pi(x) \pi(x^{\prime}).
    \label{eq:commutation}
\end{equation}

\section{Majorana Clifford operators}
\label{sec:majorana_cliffords}

To define Clifford operators in this setting, we start by introducing unitary rotations
\begin{equation}
U(\Gamma, \theta) = \exp(-i \frac{\theta}{2} \Gamma),
\end{equation}
where $\Gamma$ is a Hermitian Majorana operator and $\theta$ a real-valued angle. Because $\Gamma^2 = 1$, we have
\begin{equation}
\exp(-i \frac{\theta}{2} \Gamma) = \cos(\theta/2) - i \sin(\theta/2)\Gamma,
\end{equation}
which is same as for Pauli rotation operators. Now all other properties follow similarly, namely, for two anticommuting Majorana operators $\Gamma_1$ and $\Gamma_2$,
\begin{equation}
U(\Gamma_2, \theta)^{\dagger} \Gamma_1 U(\Gamma_2, \theta) = \cos(\theta) \Gamma_1 + i \sin(\theta)\Gamma_2 \Gamma_1.
\end{equation}
For angles $\theta = j \pi / 2$ with integer $j$, only one term on the right-hand side remains. Rotations with such angles belong to the group of Majorana Clifford operators, which transform a single Majorana operator $\Gamma$ into another Majorana operator $\Gamma^{\prime}$ instead of a general sum of Majorana operators \cite{Bettaque_Swingle:2024}. In fact, any Majorana Clifford operator can be decomposed into a product of Majorana Clifford rotations $U(\Gamma, j\frac{\pi}{2})$. In this work, we focus on a subgroup of parity-preserving Majorana Clifford operators, p-Cliffords \cite{Bettaque_Swingle:2024}, which are spanned by Majorana Clifford rotations $U(\Gamma, j\frac{\pi}{2})$ with even-parity $\Gamma$. In the remainder, we focus on the following generating set of the p-Clifford group:
\begin{equation}
\eta_{jk} = e^{\frac{\pi}{4}c_j c_k},\, \eta_j =e^{-i\frac{\pi}{4}p_j} ,\, W_{jk} = e^{-i\frac{\pi}{4}p_j p_k}.
\label{eq:majorana_clifford_generators}
\end{equation}
$\eta_j$ and $\eta_{jk}$ operators can generate arbitrary two-index Clifford rotations. By transforming $W$ operators with $\eta_{jk}$ operators, we can generate arbitrary four-index Majorana rotations. By further transforming four-index rotations with each other, we can generate rotations with strings containing 6, 8, and more Majorana operators. Finally, we note that the set (\ref{eq:majorana_clifford_generators}) is not unique and that other choices of generating operators could be considered \cite{McLauchlan2022, Mudassar_Gottesman:2024}.

\section{Phase-sensitive representation of Majorana stabilizer states}
\label{sec:stab_states}

In analogy with stabilizer states, we can define Majorana stabilizer states as
\begin{equation}
| \psi \rangle = U | 0 \rangle,
\end{equation}
where $U$ is a product of Majorana Clifford operators and $|0\rangle$ is the vacuum state. These include all computational basis states because $c_k$ and $\tilde{c}_k$ are Majorana Clifford operators.

We can represent Majorana stabilizer states by an equivalent of the phase-sensitive CH form for stabilizer states. Here, we use
\begin{equation}
    |\psi\rangle = e^{i\frac{\pi}{4}\phi} U_C U_B |s\rangle,
\end{equation}
where 
\begin{equation}
    U_C |0\rangle = |0\rangle
\end{equation}
defines a control-type Clifford operator \cite{Bravyi_Howard:2019} and $\phi$ is an integer. $U_B$ is a product
\begin{equation}
U_B = \prod_{k=0}^{n-1} B_k^{b_k}
\end{equation}
of mutually commuting braiding operators
\begin{equation}
B_k = e^{-\frac{\pi}{4} c_k \tilde{c}_{k+1}},\quad k=0,1,\dots, n-1.
\end{equation}
where $b \in \{0,1\}^{n}$.
Our algorithms will use the fact that all p-Clifford operators preserve parity to keep $b_{n-1}=0$. For convenience, we will still use the notation where $b$ has $n$ elements.

Overall, a general Majorana stabilizer state can be represented by $(\phi, U_C, b, s)$, where $U_C$ is represented by the stabilizer tableau introduced below.

\subsection{Stabilizer tableau representation of $U_C$}
\label{sec:UC_stab_tab}

$U_C$ is stored by keeping track of its action on all $c_k$ and $p_k$, in analogy with the stabilizer tableau algorithm \cite{Aaronson_Gottesman:2004, Bravyi_Howard:2019} for qubits:
\begin{eqnarray}
    U_C^{\dagger} p_j U_C &=& \prod_{k=0}^{n-1} p_k^{E_{jk}} \label{eq:tab_p} \\
    U_C^{\dagger} c_j U_C &=& i^{\omega_j} \prod_{k=0}^{n-1} p_k^{F_{jk}} c_k^{G_{jk}}, \label{eq:tab_c}
\end{eqnarray}
where $\omega_j \in \{0,1,2,3\}$ and $E, F, G$ are binary $n \times n$ matrices. In Eq.~(\ref{eq:tab_p}) we used the fact that $U_C$ is control-type, which means that it can only transform p-type operators into p-type operators without a change in phase. Indeed, let
\begin{equation}
    \Gamma = U_C^{\dagger} p_j U_C.
\end{equation}
Then,
\begin{equation}
    \Gamma |0\rangle = U_C^{\dagger} p_j U_C |0\rangle = |0\rangle,
\end{equation}
which implies that $\Gamma$ is represented by $(\varphi=0, z, x=0)$ [see Eq.~(\ref{eq:gamma_times_s})].

For an arbitrary Majorana operator $\Gamma$ described by $(\varphi, z, x)$, we can evaluate
\begin{eqnarray}
    U_C^{\dagger} \Gamma U_C
    &=& U_C^{\dagger} i^{\varphi} \prod_j p_j^{z_j} c_j^{x_j} U_C \\
    &=& i^{\varphi} \prod_j U_C^{\dagger} p_j^{z_j} U_C U_C^{\dagger} c_j^{x_j} U_C \\
    &=& i^{\varphi} \prod_j \prod_{k=0}^{n-1} p_k^{E_{jk}z_j} i^{\omega_jx_j} \prod_{k=0}^{n-1} p_k^{F_{jk}x_j} c_k^{G_{jk}x_j} \\
    &=& i^{\varphi} \prod_j \Gamma^{(j)}, \\
    \Gamma^{(j)} &=& i^{\omega_jx_j} \prod_{k=0}^{n-1} p_k^{E_{jk}z_j + F_{jk}x_j} c_k^{G_{jk}x_j}
\end{eqnarray}
This operation scales as $\mathcal{O}((|z| + |x|) n)=\mathcal{O}(n^2)$ for arbitrary Majorana operators.

Due to commutation relations between $c_j$ and $p_j$ Majorana operators, matrices $E$, $F$, and $G$ are not completely independent. In Appendix B, we derive the following identities:
\begin{eqnarray}
    E \cdot G^{T} = E^T \cdot G  &=& I, \label{eq:main_tableau_identity1} \\
    F \cdot G^T + G \cdot F^{T} &=& I + G \cdot G^T. \label{eq:main_tableau_identity2}
\end{eqnarray}
Furthermore, each row of matrix $G$ must have an odd number of ones because it corresponds to the transformation of an odd-parity, local Majorana operator $c_j$.

\subsection{Transformation of a Majorana operator under $U_B$}
\label{sec:majorana_trans_UB}

A key ingredient in many procedures is the conjugation of a general Majorana operator $\Gamma$ by $U_B$,
\begin{equation}
    \Gamma' = U_B^{\dagger} \Gamma U_B,
\end{equation}
where we assume $U_B$ is provided as a binary array $b$ and $\Gamma$ by $(\varphi, z, x)$. To evaluate $\Gamma'$, we first need to check which $B_k$ do not commute with $\Gamma$:
\begin{equation}
    \beta_k = B_k^{b_k} \, \text{does not commute with}\, \Gamma = b_k \cdot (x_k + z_k + z_{k+1}).
\end{equation}
For each $B_k^{b_k}$ that does not commute with $\Gamma$, we will have a factor of $c_k\tilde{c}_{k+1}$ multiplying $\Gamma$ from the left. This is because:
\begin{equation}
    e^{\frac{\pi}{4}c_k\tilde{c}_{k+1}} \Gamma_C e^{-\frac{\pi}{4}c_k\tilde{c}_{k+1}} =
    e^{\frac{\pi}{2}c_k\tilde{c}_{k+1}} \Gamma_C = c_k \tilde{c}_{k+1} \Gamma_C,
\end{equation}
and because all $B_k$ commute with each other. Finally, we need to compute the product 
\begin{equation}
    \Gamma' = \Gamma_B (\beta) \Gamma,
\end{equation}
where
\begin{equation}
    \Gamma_B (\beta) = \prod_{k=0}^{n-1} (c_k\tilde{c}_{k+1})^{\beta_k} = \prod_{k=0}^{n-1} (-ic_k p_{k+1} c_{k+1})^{\beta_k} = i^{-|\beta|} c_0^{\beta_0} \prod_{k=1}^{n-1} p_k^{\beta_{k-1}} c_k^{\beta_{k-1} + \beta_k}.
    \label{eq:gamma_B}
\end{equation}
In Eq.~(\ref{eq:gamma_B}), we used the fact that $b_{n-1}=0$ (and hence $\beta_{n-1}=0$).

\subsection{Expectation value of a Majorana operator}
\label{sec:majorana_exp_val}

Given a Majorana stabilizer state $|\psi\rangle$ and a Majorana operator $\Gamma$, we can compute the expectation value
\begin{equation}
    \langle\psi|\Gamma|\psi\rangle = i^{\varphi^{\prime}} (-1)^{z^{\prime} \cdot s} \delta_{x^{\prime}},
\end{equation}
where $(\varphi^{\prime}, z^{\prime}, x^{\prime})$ correspond to $\Gamma^{\prime} = U_B^{\dagger}U_C^{\dagger} \Gamma U_C U_B$ and
\begin{equation}
    \delta_x = \begin{cases}
        1, \quad x_0 = x_1 = \cdots = x_{n-1} =0, \\
        0, \quad \text{otherwise}.
    \end{cases}
\end{equation}
$\Gamma'$ can be computed using procedures introduced in Secs.~\ref{sec:UC_stab_tab} and \ref{sec:majorana_trans_UB}.

\subsection{Action of Majorana Clifford operators on Majorana stabilizer states (summary)}

The key feature of this representation of Majorana stabilizer states is that we can efficiently apply Majorana Clifford operators to them. As discussed in Sec.~\ref{sec:majorana_cliffords}, we only need to derive the update rules for acting by $\eta_j$, $W_{jk}$, and $\eta_{jk}$. In addition, we implemented the action of a general Majorana operator and Majorana Clifford rotations on a Majorana stabilizer state. A summary is given in Table~\ref{tab:summary}, while the explicit update rules are given in Appendix A. We note here that $\eta_j$ and $W_{jk}$ are diagonal in the computational basis, which means that we only need to update the phase and $U_C$. In contrast, a general Majorana operator, Majorana Clifford rotation, and $\eta_{jk}$ can, in principle, 
affect all parameters of the Majorana stabilizer state. The update rules due to $\eta_{jk}$ are particularly involved and the algorithm follows basic ideas from the update rules for applying Hadamard gates to the stabilizer states in the CH-form.

\begin{table}[ptb]
    \centering
    \caption{Summary of results for our phase-sensitive representation of Majorana stabilizer states. The update rules for applying various fermionic gates are given in Appendix~A, whereas the computation of amplitudes and inner products is given in Secs.~\ref{sec:amplitudes} and \ref{sec:inner_prod}.}
    \begin{tabular}{l|c}
        \hline \hline
       Operation  & Complexity \\ \hline
        Apply $\eta_j$ or $W_{jk}$ & $\mathcal{O}(n)$ \\
        Apply $\eta_{jk}$ & $\mathcal{O}(n^2)$ \\
        Apply general Majorana Clifford rotation $\exp(-i \frac{\pi}{4} \Gamma )$ & $\mathcal{O}(n^2)$ \\
        Apply general Majorana operator $\Gamma$ & $\mathcal{O}(n^2)$ \\
        Probability amplitude $\langle x | \psi \rangle$ & $\mathcal{O}((|x| + c) n)$ \\
        Inner product $\langle \phi | \psi \rangle$ & $\mathcal{O}(n^3)$ \\
         \hline \hline
    \end{tabular}
    \label{tab:summary}
\end{table}

\subsection{Probability amplitudes of Majorana stabilizer states}
\label{sec:amplitudes}
The first step in computing probability amplitudes of the form $\langle x | \psi\rangle$, where $|\psi\rangle = e^{i\pi \phi/4} U_C U_B |s\rangle$ and $x$ a computational basis state, is to write
\begin{equation}
    \langle x| = \langle 0 | \Gamma, \quad \Gamma = (-1)^{\lfloor |x|/2 \rfloor} \prod_{k=0}^{n-1} c_k^{x_k}.
\end{equation}
We can apply $\Gamma$ to $|\psi\rangle$ to obtain updated $\phi$, $U_C$, $b$, and $s$. Therefore, we now only need a procedure to compute $\langle 0 | \psi\rangle$, which can be done in $\mathcal{O}(|x|n)$ time. We first note that this amplitude is zero if $s$ is of odd parity, $\pi(s) = 1$. If $\pi(s)=0$, we have
\begin{eqnarray}
    \langle 0 | e^{i\frac{\pi}{4}\phi} U_C U_B |s\rangle 
    &=& e^{i\frac{\pi}{4}\phi} \langle 0 | U_B |s\rangle \\
    &=& 2^{-|b|/2} e^{i\frac{\pi}{4}\phi} \langle 0 | \prod_{k=0}^{n-1} (1 - c_k \tilde{c}_{k+1})^{b_k} |s\rangle \label{eq:amplitude_der_1} \\
    &=& 2^{-|b|/2} e^{i\frac{\pi}{4}\phi} \langle 0 | \prod_{k=0}^{n-1} (-c_k \tilde{c}_{k+1})^{\bar{s}_k} |s\rangle \delta_{b \bar{s},\bar{s}} \label{eq:amplitude_der_2} \\
    &=& 2^{-|b|/2} e^{i\frac{\pi}{4}\phi} i^{|\bar{s}|} \langle 0 | \prod_{k=0}^{n-1} p_k^{\bar{s}_{k-1}} c_k^{s_k} |s\rangle \delta_{b \bar{s},\bar{s}} \label{eq:amplitude_der_3} \\
    &=& 2^{-|b|/2} e^{i\frac{\pi}{4}\phi} i^{|\bar{s}|} (-1)^{|s|/2} \langle 0 | \prod_{k=0}^{n-1} p_k^{\bar{s}_{k-1}} |0\rangle \delta_{b \bar{s},\bar{s}} \\
    &=& 2^{-|b|/2} e^{i\frac{\pi}{4}\phi} i^{|\bar{s}|} (-1)^{|s|/2} \delta_{b \bar{s},\bar{s}} \\
    &=& 2^{-|b|/2} e^{i\frac{\pi}{4}\phi} i^{|\bar{s}| + |s|} \delta_{b \bar{s},\bar{s}}
\end{eqnarray}
In going from (\ref{eq:amplitude_der_1}) to (\ref{eq:amplitude_der_2}), we reduced the exponential sum of Majorana operators to a single Majorana operator that has a non-zero off-diagonal matrix element between the $|0\rangle$ and $|s\rangle$ states. This Majorana operator is exactly the product
\begin{equation}
    (-c_k\tilde{c}_{k+1})^{\bar{s}_k} = i^{|\bar{s}|} \prod_{k=0}^{n-1} p_k^{\bar{s}_{k-1}} c_k^{s_k},
\end{equation}
and is present in Eq.~(\ref{eq:amplitude_der_1}) only if $b_k=1$ for every $k$ at which $\bar{s}_k=1$. This condition is enforced by the delta function:
\begin{equation}
    \delta_{b\bar{s}, \bar{s}} = \begin{cases}
        1, \quad b_k \bar{s}_k = \bar{s}_k, \quad k=0, 1, \dots, n-1, \\
        0, \quad \text{otherwise}.
    \end{cases}
\end{equation}

\subsection{Inner product of two Majorana stabilizer states}
\label{sec:inner_prod}
Given stabilizer states $|\psi\rangle$ and $|\psi'\rangle$, we can compute their inner product in $\mathcal{O}(n^3)$ time.
\begin{eqnarray}
    \langle \psi' | \psi \rangle 
    &=& e^{-i\frac{\pi}{4}\phi'} \langle s' | U_B'^{\dagger} U_C'^{\dagger} | \psi\rangle \\
    &=& e^{-i\frac{\pi}{4}\phi'} \langle s' | U_C'^{\dagger} \prod_{j=0}^{n-1} e^{-i b_j' \frac{\pi}{4}\Gamma_j} | \psi\rangle \\
    &=& e^{-i\frac{\pi}{4}\phi'} \langle s' | U_C'^{\dagger} | \bar{\psi}\rangle \\
    &=& e^{-i\frac{\pi}{4}\phi'} (-i)^{\theta} \langle t | \bar{\psi}\rangle,
    \label{eq:inner_product}
\end{eqnarray}
 where we introduced
 \begin{eqnarray}
     \Gamma_j &=& U_C' i c_j \tilde{c}_{j+1} U_C'^{\dagger} \label{eq:gamma_j_inner_product}, \\
     U_C' |s'\rangle &=& i^{\theta} |t\rangle,
     \label{eq:UCprime_inner_product}
 \end{eqnarray}
 and
 \begin{equation}
     |\bar{\psi}\rangle = \prod_{j=0}^{n-1} e^{-i b_j' \frac{\pi}{4}\Gamma_j} | \psi\rangle.
 \end{equation}
 Equation~(\ref{eq:inner_product}) can be evaluated using expressions derived in Sec.~\ref{sec:amplitudes} for the amplitudes of stabilizer states. Two additional subroutines needed to evaluate Eqs.~(\ref{eq:gamma_j_inner_product}) and (\ref{eq:UCprime_inner_product}) are the construction of the stabilizer tableaux for $U_C^{\dagger}$ and applying $U_C$ on a computational basis state. These are given in Appendix B.

 The main source of cubic scaling of the inner product is the fact that constructing stabilizer tableux for $U_C^{\dagger}$ involves matrix products. Another source is the fact that there are $\mathcal{O}(n)$ nonzero elements of $b'$ and for each of these we have to apply one general Majorana Clifford rotation with a cost of $\mathcal{O}(n^2)$. As a minor improvement to the algorithm, we first check if $b<b'$, in which case we compute $\langle \psi | \psi' \rangle$ and obtain the final result as its complex conjugate. The cost of this part of the computation is then $\mathcal{O}(\min(|b|, |b'|) n^2)$.

 \section{Conclusion}

 We have presented a set of algorithms for representing Majorana stabilizer states, updating them upon action by parity-preserving Majorana Clifford operators, and computing probability amplitudes and inner products. We also provide a basic Python code at \url{github.com/tbegusic/fermionic-stabilizer-states} that implements these functions and can be used to verify the correctness of the algorithms by comparing against Qiskit \cite{javadiabhari2024quantumcomputingqiskit} Statevector simulator running in the Jordan-Wigner representation.

 Overall, these algorithms are needed for representing general fermionic states by superpositions of Majorana stabilizer states, $|\psi\rangle = \sum_{i} \alpha_i |\phi_i\rangle$. Specifically, they allow us to prepare Majorana stabilizer states, evaluate overlaps $\langle \phi_i | \phi_j \rangle$, and Hamiltonian (or other operator) matrix elements $\langle \phi_i | H | \phi_j\rangle$. Regarding the simulation of fermionic quantum circuits, our algorithms enable low-rank stabilizer methods working directly in the fermionic picture.

 \begin{acknowledgments}
 The authors were supported by the U.S. Department
of Energy, Office of Science, Accelerated Research in Quantum Computing Centers, Quantum
Utility through Advanced Computational Quantum Algorithms, grant no. DE-SC0025572. TB acknowledges partial financial support from the Swiss National Science Foundation through the Postdoc Mobility Fellowship (grant number P500PN-214214). GKC is a Simons Investigator in Physics.
 \end{acknowledgments}
 
\appendix

\section{Majorana stabilizer state procedures}

\subsection{Update rules for applying a Majorana operator}

The procedure to evaluate $|\psi^{\prime} \rangle = \Gamma | \psi\rangle$ is closely related to the computation of the expectation value of a Majorana operator (Sec.~\ref{sec:majorana_exp_val}). Let us assume we are given $\Gamma$ through $(\varphi, z, x)$ and $| \psi \rangle$ through $(\phi, U_C, b, s)$. Then,
\begin{eqnarray}
    |\psi^{\prime} \rangle &=& \Gamma | \psi\rangle \\
    &=& e^{i\frac{\pi}{4}\phi} U_C U_B \Gamma^{\prime} |s\rangle \\
    &=& e^{i\frac{\pi}{4}\phi} i^{\varphi^{\prime}} (-1)^{(s+x^{\prime})\cdot z^{\prime} + s \cdot \bar{x}^{\prime} + \pi(x^{\prime})\pi(s)} U_C U_B  |s+x^{\prime}\rangle,
\end{eqnarray}
where $(\varphi^{}\prime, z^{\prime}, x^{\prime})$ are the parameters of $\Gamma^{\prime} = U_B^{\dagger} U_C^{\dagger} \Gamma U_C U_B$. The update rules are:
\begin{eqnarray}
    \phi &\leftarrow& \phi + 2\varphi' + 4[(s+x^{\prime})\cdot z^{\prime} + s \cdot \bar{x}^{\prime} + \pi(x^{\prime})\pi(s)] \\
    s &\leftarrow& s + x'
\end{eqnarray}

\subsection{Update rules for applying $\eta_j$ and $W_{jk}$}

These gates are diagonal in the computational basis and are control-type up to a phase. As shown below, this means that we only need to update $\phi$ and $U_C$.

Let us first consider $\eta_{j}=\exp(-i\pi p_j / 4)$ and a stabilizer state defined by $(\phi, U_C, b, s)$, where $U_C$ is represented by a set of parameters $(\omega, E, F, G)$ as in Sec.~\ref{sec:UC_stab_tab}. We have
\begin{equation}
    \eta_{j} |\psi\rangle = e^{-i \frac{\pi}{4}} e^{i \frac{\pi}{4}(1- p_j)} e^{i \frac{\pi}{4}\phi} U_C U_B |s\rangle = e^{i \frac{\pi}{4}(\phi-1)} U_C^{\prime} U_B |s\rangle,
\end{equation}
where $U_C^{\prime} = e^{i \frac{\pi}{4}(1- p_j)} U_C$ and $e^{i \frac{\pi}{4}(1- p_j)}$ is a control-type operator satisfying
\begin{equation}
    e^{i \frac{\pi}{4}(1- p_j)} |0\rangle = |0\rangle.
\end{equation}
We now need to derive the update rules for $\omega$, $E$, $F$, and $G$ to represent $U_C^{\prime}$.

$E$ is not modified because
\begin{equation}
    U_C^{\dagger} \eta_j^{\dagger} p_k \eta_j U_C = U_C^{\dagger} p_k U_C.
\end{equation}
For the remaining parameters, we evaluate
\begin{eqnarray}
   U_C^{\dagger} \eta_j^{\dagger} c_j \eta_j U_C 
   &=& U_C^{\dagger} i p_j c_j U_C \\
   &=& i \prod_{k=0}^{n-1} p_k^{E_{j,k}} i^{\omega_j} \prod_{k=0}^{n-1} p_k^{F_{j,k}} c_k^{G_{j,k}} \\
   &=& i^{\omega_j+1} \prod_{k=0}^{n-1} p_k^{E_{j,k}+F_{j,k}} c_k^{G_{j,k}}
\end{eqnarray}
and
\begin{equation}
    U_C^{\dagger} \eta_j^{\dagger} c_k \eta_j U_C = U_C^{\dagger} c_k U_C, \quad k \neq j.
\end{equation}
Therefore, the update rules are
\begin{equation}
    \phi \leftarrow \phi - 1, \quad F_j \leftarrow F_j + E_j, \quad \omega_j \leftarrow \omega_j + 1.
\end{equation}
In the code provided with this manuscript, we implemented $\eta_j^{\pm} = \exp(\mp i \frac{\pi}{4} p_j)$, for which the update rules are simply
\begin{equation}
    \phi \leftarrow \phi \mp 1, \quad F_j \leftarrow F_j + E_j, \quad \omega_j \leftarrow \omega_j \pm 1.
\end{equation}

Similarly, for $W_{jk}$, we have
\begin{eqnarray}
   U_C^{\dagger} W_{jk}^{\dagger} c_j W_{jk} U_C 
   &=& U_C^{\dagger} i p_j p_k c_j U_C \\
   &=& i \prod_{l=0}^{n-1} p_l^{E_{j,l}} \prod_{l=0}^{n-1} p_l^{E_{k,l}} i^{\omega_j} \prod_{l=0}^{n-1} p_l^{F_{j,l}} c_l^{G_{j,l}} \\
   &=& i^{\omega_j+1} \prod_{l=0}^{n-1} p_l^{E_{j,l}+ E_{k,l} + F_{j,l}} c_k^{G_{j,k}}
\end{eqnarray}
and the update rules are
\begin{eqnarray}
    \phi &\leftarrow& \phi - 1, \\
    \quad F_{j/k} &\leftarrow& F_{j/k} + E_j + E_k, \quad \omega_{j/k} \leftarrow \omega_{j/k} + 1.
\end{eqnarray}
As for $\eta_j$, here we also implemented a more general version $W_{jk}^{\pm} = \exp(\mp i \frac{\pi}{4} p_j p_k)$ with update rules
\begin{eqnarray}
    \phi &\leftarrow& \phi \mp 1, \\
    \quad F_{j/k} &\leftarrow& F_{j/k} + E_j + E_k, \quad \omega_{j/k} \leftarrow \omega_{j/k} \pm 1.
\end{eqnarray}

\subsection{Update rules for applying $\eta_{jk}$}

The most complicated part is the update of the Majorana stabilizer state after applying $\eta_{jk}$ ($j \neq k$). Following the strategy for applying the Hadamard gate to the CH form for qubit stabilizer states, the basic steps to derive the algorithm involve:
\begin{enumerate}
\item Introduce $\Gamma = U_B^{\dagger} U_C^{\dagger} c_j c_k U_C U_B$, so that
\begin{equation}
\eta_{jk} U_C U_B |s \rangle = U_C U_B \frac{1}{\sqrt{2}} (1 + \Gamma) |s \rangle = U_C U_B \frac{1}{\sqrt{2}}(|s\rangle + i^{\theta}|t\rangle).
\label{eq:step1}
\end{equation}
\item Find Majorana Clifford operator $C$, such that
\begin{equation}
C(|s\rangle + i^{\theta}|t\rangle) = |s'\rangle + i^{\theta'}|t'\rangle, s_w^{\prime} \neq t_w^{\prime}, s_{w+1}^{\prime} \neq t_{w+1}^{\prime}, s_{k\neq w,w+1}^{\prime} = t_{k\neq w,w+1}^{\prime},
\end{equation}
i.e., that $s'$ and $t'$ differ only at two neighboring sites $w$ and $w+1$, and that
\begin{equation}
U_B C^{\dagger} U_B^{\dagger} = V_C,
\end{equation}
where $V_C$ is a control-type Clifford. This means that we can rewrite
\begin{equation}
U_C U_B \frac{1}{\sqrt{2}}(|s\rangle + i^{\theta}|t\rangle) = U_C V_C U_B \frac{1}{\sqrt{2}}(|s'\rangle + i^{\theta'}|t'\rangle)
\label{eq:step2}
\end{equation}
and simply update $U_C \leftarrow U_C V_C$.
\item Rewrite $\exp(i \pi \phi/4)U_C U_B (|s'\rangle + i^{\theta'}|t'\rangle)$ into $\exp(i \pi \phi^{\prime}/4) U_C^{\prime} U_B^{\prime} |u \rangle$.
\end{enumerate}
The first step can be performed in $O(n)$ time because the initial Majorana string $c_j c_k$ is local. In the second step, $V_C$ will be a non-local operator and updating the $E, F, G$ matrices will lead to $O(n^2)$ runtime. The third step can be performed in $O(n^2)$ time because in certain cases we need to update $U_C$ with a non-local control-type operator. The overall runtime is $O(n^2)$, which coincides with the standard CH update algorithm for a Hadamard gate. Let us now look into each of these steps in detail.

\subsubsection*{Step 1}
We are given $\eta_{jk}$ (in the form of sites $j$ and $k$) and a stabilizer state with parameters $(\phi, \omega, E, F, G, b, s)$. We first compute the parameters $(\varphi, z, x)$ of $\Gamma = U_B^{\dagger} U_C^{\dagger} c_j c_k U_C U_B$ as described earlier. We can evaluate $\theta$ and $t$ of Eq.~(\ref{eq:step1}) by using Eq.~(\ref{eq:gamma_times_s}). In the following Section, we will use $x=s+t$ to denote a binary array with 1 where $s$ and $t$ are different and 0 where they are equal.

\subsubsection*{Step 2}
Our goal now is to modify $s$ and $t$ so that they differ only at two neighboring sites. To this end, we introduce a control operator
\begin{equation}
    C_{\Gamma_1, \Gamma_2} = e^{-i\frac{\pi}{4}(I-\Gamma_1)(I-\Gamma_2)},
\end{equation}
where $[\Gamma_1, \Gamma_2]=0$ and both $\Gamma_1$ and $\Gamma_2$ are Hermitian Majorana operators. $C_{\Gamma_1, \Gamma_2}^2 = I$ and therefore $C_{\Gamma_1, \Gamma_2}$ is Hermitian. We further impose that one of the two operators is a p-type Majorana operator, for example, $\Gamma_1=\prod_{k \in K} p_k$ for some set of indices $K$. Then $C_{\Gamma_1, \Gamma_2}$ is a control-type operator in the sense that $C|0\rangle = |0\rangle$. It has the property that
\begin{equation}
    C_{\Gamma_1, \Gamma_2} |s\rangle = \begin{cases}
        |s\rangle, \quad \Gamma_1 |s\rangle = |s\rangle \\
        \Gamma_2 |s\rangle, \quad \Gamma_1 |s\rangle = -|s\rangle.
    \end{cases}
\end{equation}
Below, we identify $\Gamma_1$ and $\Gamma_2$ that allow us to implement the desired transformation of $|s\rangle$ and $|t\rangle$.

We begin by defining an array of indices $v$, which stores all indices $i$ at which $b_i=0$. Let $n_v$ denote the number of elements in $v$ (i.e., the number of zeros in $b$). We assume that $b_{n-1}=0$, so $n-1$ is always the last element of $v$ and $n_v>0$. We can also define $n_v$ arrays of indices $S_0=[0, v_0]=(0, 1, \dots, v_0)$ and $S_{0<j<n_v} = (v_{j-1}, v_{j}] = (v_{j-1}+1, v_{j-1}+2, \dots, v_j)$. Now the operators of the form $\prod_{k \in S_j}p_k$ commute with $U_B$ because they anticommute only with $B_{v_{j-1}}$ (for $j>0$) and $B_{v_j}$, which are not present in $U_B$ by definition of $v$.

We must distinguish two cases:

1. There exists $j \in [0, n_v)$ such that $\sum_{k \in S_j} x_k = 1$.

2. $\sum_{k \in S_j} x_k = 0$ for all $j \in [0, n_v)$.

In words, we ask if there is any subset of indices $S_j$ on which $s$ and $t$ differ odd number of times.

\textit{Case 1.} Let us assume there exists $j$ such that $\sum_{k \in S_j} x_k = 1$. We choose the smallest value of $j$ and set
\begin{equation}
    \Gamma_1 = \prod_{k \in S_j} p_k.
\end{equation}
If $\sum_{k\in S_j} s_k = 0$, then $\Gamma_1|s\rangle = |s\rangle$ and, because we know that $s$ and $t$ differ an odd number of times on $S_j$ indices, $\Gamma_1|t\rangle = -|t\rangle$. Otherwise, if $\sum_{k\in S_j} s_k = 1$, we can rewrite
\begin{equation}
    |s\rangle + i^{\theta} |t\rangle = i^{\theta} (|t\rangle + i^{-\theta} |s\rangle),
\end{equation}
which means that after substituting $s \leftarrow t$, $t \leftarrow s$, $\phi \leftarrow \phi + 2\theta$, $\theta \leftarrow -\theta$, we again recover $\Gamma_1|s\rangle = |s\rangle$ and $\Gamma_1|t\rangle = -|t\rangle$.
Therefore,
\begin{equation}
    C_{\Gamma_1, \Gamma_2} |s\rangle = |s\rangle, \quad 
    C_{\Gamma_1, \Gamma_2} | t \rangle = \Gamma_2 |t\rangle.
\end{equation}
We set $\Gamma_2$ as
\begin{eqnarray}
    \Gamma_2 &=&  i^{|x^{\prime}|/2} \prod_{k=0}^{n-1} c_k^{x^{\prime}_k}, \\
    x^{\prime} &=& x + e_w + e_{w+1} \\
    w &=&
    \begin{cases}
        v_{j-1}, \quad j>0 \\
        v_0,  \quad j=0,
        \label{eq:w}
     \end{cases}
\end{eqnarray}
where $|x|$ denotes the number of nonzero entries in $x$, $e_w$ is a unit vector with 1 at position $w$ and zeros elsewhere, and the prefactor ensures that $\Gamma_2$ is Hermitian.

We now have
\begin{equation}
     U_C U_B (|s\rangle + i^{\theta} |t\rangle)=
    U_C U_B C_{\Gamma_1, \Gamma_2} C_{\Gamma_1, \Gamma_2} (|s\rangle + i^{\theta} |t\rangle) = U_C C_{\Gamma_1, \Gamma_2^{\prime}} U_B (|s\rangle + i^{\theta^{\prime}} |t^{\prime}\rangle),
\end{equation}
where
\begin{eqnarray}
    \Gamma_2^{\prime} &=& U_B \Gamma_2 U_B^{\dagger}, \\
    i^{\theta^{\prime}} |t^{\prime}\rangle &=& i^{\theta} \Gamma_2 |t\rangle,
\end{eqnarray}
and, by construction, $|t^{\prime}\rangle = |s+e_w + e_{w+1}\rangle$. 

Note that $w \neq n-1$, which can be seen easily from Eq.~(\ref{eq:w}): It is not possible in the first case and in the second case it would imply $v_0=n-1$, which would mean that $s$ and $t$ differ in total on an odd number of sites (this is not possible because we consider only parity-preserving Majorana Clifford gates).

\textit{Case 2.} Let us assume that $\sum_{k\in S_j}x_k=0$ for all $j\in [0,n_v)$. Now we cannot choose $\Gamma_1$ such that it commutes with $U_B$ and $\Gamma_1 |s\rangle = - \Gamma_1 |t\rangle$. We therefore choose 
\begin{equation}
 \Gamma_2 = (-1)^{s_w} p_w ,  
\end{equation}
where $w$ is the first site at which $s_w \neq t_w$. Then $\Gamma_2|s\rangle = |s\rangle$ and $\Gamma_2|t\rangle = -|t\rangle$. Our strategy will be to identify $\Gamma_1$ such that
\begin{equation}
    \Gamma_1^{\prime} = U_B \Gamma_1 U_B^{\dagger} = \prod_{k \in K} p_k, \quad [\Gamma_1, \Gamma_2] = 0,
    \label{eq:gamma_2_conditions}
\end{equation}
and that $|t^{\prime}\rangle \propto \Gamma_1|t\rangle$ and $|s\rangle$ differ only at two neighboring sites. Indeed, in that case we would have
\begin{equation}
    U_C U_B (|s\rangle + i^{\theta} |t\rangle
    = U_C U_B C_{\Gamma_1, \Gamma_2}^{\dagger} C_{\Gamma_1, \Gamma_2} (|s\rangle + i^{\theta} |t\rangle)
    = U_C C_{\Gamma_1^{\prime}, \Gamma_2^{\prime}}^{\dagger} U_B (|s\rangle + i^{\theta^{\prime}} |t^{\prime}\rangle),
\end{equation}
where $C_{\Gamma_1^{\prime}, \Gamma_2^{\prime}}$ is a control-type operator because of Eq.~(\ref{eq:gamma_2_conditions}), i.e., $C_{\Gamma_1^{\prime}, \Gamma_2^{\prime}}|0\rangle = |0\rangle$, and can be merged into the representation of $U_C$ by updating its stabilizer tableau representation. What follows is the construction of $\Gamma_1$.

Let us define Majorana operators
\begin{equation}
    Y_j = i c_j c_{j+1} \prod_{k=j+2}^{n-1} p_k, \quad j \in [0,n-1],
\end{equation}
which satisfy
\begin{equation}
    B_j Y_j B_j^{\dagger} = \prod_{k=j+1}^{n-1}p_k \quad \text{and} \quad [Y_j, B_{i \neq j}]=0.
\end{equation}
We also introduce an array $m$ containing indices of nonzero elements in the binary array $x$ (the difference of $s$ and $t$). By assumption of \textit{Case 2}, we know that all $b_{m_{2k} \leq j < m_{2k+1}} = 1$. Otherwise, if any of these were $0$, we would have odd number of nonzero elements of $x$ in $S_j$, which would correspond to \textit{Case 1}. This motivates us to define
\begin{equation}
    \Lambda_{i_1, i_2} = \prod_{j=i_1}^{i_2-1}Y_j, \quad m_{2k} \leq i_1 \leq i_2 \leq m_{2k+1} \quad k=[0,|x|/2)
    \label{eq:lambda}
\end{equation}
that can be written out explicitly as
\begin{eqnarray}
    \Lambda_{i_1, i_2} &=&
    \begin{cases}
        (-1)^{(i_2-i_1)/2} c_{i_1} p_{i_1+2} p_{i_1+4} \dots p_{i_2} c_{i_2}, \quad i_2-i_1 = \,\text{even},\\
        (-1)^{(i_2-i_1-1)/2} i c_{i_1} p_{i_1+2} p_{i_1+4} \dots p_{i_2-1} c_{i_2} \prod_{j=i_2+1}^{n-1} p_j, \quad i_2-i_1 =\, \text{odd}\\
    \end{cases}\\
    &=& i^{i_2-i_1} c_{i_1} \prod_{k=1}^{\lfloor \frac{i_2-i_1}{2} \rfloor} p_{i_1+2k} c_{i_2} \prod_{j=i_2+1}^{n-1} p_j^{(i_2-i_1) \,\text{mod}\, 2}
    \label{eq:lambda_explicit}
\end{eqnarray}
and satisfy
\begin{eqnarray}
    U_B \Lambda_{i_1, i_2} U_B^{\dagger}
    &=& \prod_{j=i_1}^{i_2-1} B_j Y_j B_j^{\dagger} \\
    &=& \prod_{j=i_1}^{i_2-1} \prod_{l=j+1}^{n-1}p_l \\
    &=& \begin{cases}
        p_{i_1+1} p_{i_1+3} \dots p_{i_2-1}, \quad i_2-i_1 = \,\text{even},\\
        p_{i_1+1} p_{i_1+3} \dots p_{i_2-2} \prod_{j=i_2}^{n-1}p_j, \quad i_2-i_1 =\, \text{odd}.\\
    \end{cases}\\
    &=& \prod_{k=0}^{\lfloor \frac{i_2 - i_1-1}{2} \rfloor} p_{i_1 + 1 + 2k} \prod_{j=i_2+1}^{n-1} p_j^{(i_2-i_1) \,\text{mod}\, 2}
    \label{eq:lambda_transformed}
\end{eqnarray}

We can see from Eq.~(\ref{eq:lambda_explicit}) that $\Lambda_{i_1, i_2}$ changes a computational basis state at exactly two sites ($i_1$ and $i_2$), while from Eq.~(\ref{eq:lambda_transformed}) we see that these operators transform under $U_B$ into a product of diagonal operators [see (\ref{eq:gamma_2_conditions})]. Therefore, we can use them to construct
\begin{equation}
    \Gamma_1 = \Lambda_{m_0+1, m_1} \prod_{k=1}^{|x|/2-1} \Lambda_{m_{2k}, m_{2k+1}},
    \label{eq:gamma_1_case2}
\end{equation}
which commutes with $\Gamma_2$, satisfies (\ref{eq:gamma_2_conditions}), and
\begin{equation}
    \Gamma_1 i^{\theta} |t \rangle = i^{\theta^{\prime}} |s + e_w + e_{w+1}\rangle, \quad w=m_0.
\end{equation}

As in \textit{Case 1}, $w \neq n-1$ because $m_0=n-1$ would imply that $s$ and $t$ differ only at one site.

$\Gamma_1$ can be constructed in $\mathcal{O}(n)$ time by preparing separately the indices enclosed by $i_1$ and $i_2$ [first part of Eq.~(\ref{eq:lambda_explicit})] and the tails composed of $p_j$ operators [last part of Eq.~(\ref{eq:lambda_explicit})]. For the former, we simply iterate over all pairs of $m_{2k}$ and $m_{2k+1}$ and fill in values for $x$ and $z$. From (\ref{eq:lambda_explicit}) and (\ref{eq:gamma_1_case2}), the phase is
\begin{equation}
    \varphi = -1 + \sum_{k=0}^{|x|/2-1} (m_{2k+1} - m_{2k}).
\end{equation}
For the tails of $p_j$ operators, we have
\begin{equation}
    z^{(P)}_{m_{2k-1}+1 \leq j \leq m_{2k+1}} = (\sum_{j=0}^{2k+1} m) \,\text{mod}\, 2, \quad k=1, 2, \dots, |x|/2-1.
\end{equation}
To derive this expression, we used
\begin{equation}
    [(m_1 - m_0) + (m_3 - m_2) + \dots + (m_{2k+1}-m_{2k})] \,\text{mod}\, 2 = (\sum_{j=0}^{2k+1} m_j) \,\text{mod}\, 2.
\end{equation}
We note that these $p$ operators are already ordered within $\Gamma_1$, so to include them, we simply update $z$ prepared above by
\begin{equation}
    z \leftarrow z + z^{(P)}.
\end{equation}

Finally, to complete Step 2, we need to show how to update $(\omega, E, F ,G)$ to represent $U_C^{\prime} = U_C C_{\Gamma_1, \Gamma_2}$, where $\Gamma_1$ is a product of $p_k$ and $\Gamma_2$ is an arbitrary Hermitian Majorana operator. This is given in Appendix B.

\subsubsection*{Step 3}

Let us now assume that we have
\begin{equation}
|\psi\rangle = e^{i\frac{\pi}{4}\phi}U_C U_B \frac{1}{\sqrt{2}}(|s\rangle + i^{\theta} |s+e_w+e_{w+1}\rangle),
\end{equation}
in the form of stabilizer tableau representation for $U_C$, binary arrays $b$ (for $U_B$) and $s$, integer $\theta \in \{0,1,2,3\}$, and integer index $w$.
We have two cases:

\textit{Case 1.} $\theta = 2k+1$ ($k=0,1$) is odd.
Then,
\begin{eqnarray}
    B_w^{b_w} \frac{1}{\sqrt{2}}(|s\rangle + i^{\theta} |s+e_w+e_{w+1}\rangle) 
    &=& B_w^{b_w+1} (c_w\tilde{c}_{w+1})^{s_w+s_{w+1}+k}|s\rangle \\ 
    &=& \begin{cases}
        B_w^{b_w+1} |s\rangle, \quad s_w+s_{w+1}+k + b_w=1 \\
        B_w^{b_w+1} i^{\theta} |s+e_w+e_{w+1}\rangle, \quad s_w+s_{w+1}+k+b_w=0
    \end{cases},
\end{eqnarray}

Therefore, we only need to update $b \leftarrow b+e_w$ and, if $s_w + s_{w+1} + k + b_w = 1$, update phase $\phi \leftarrow \phi  + 2\theta$ and bitstring $s \leftarrow s + e_w + e_{w+1}$.

\textit{Case 2.} $\theta = 2k$ ($k=0,1$) is even.

We first show that
\begin{equation}
    B_w^{b_w} \frac{1}{\sqrt{2}}(|s\rangle + i^{\theta} |s+e_w+e_{w+1}\rangle)
    = e^{-i \frac{\pi}{4}b_w(-1)^{s_w + s_{w+1} + k}} e^{-i\frac{\pi}{4}(1-\Gamma)}\frac{1}{\sqrt{2}} (|s\rangle + i^{\theta+1} |s+e_w+e_{w+1}\rangle),
\end{equation}
where $\Gamma = (-1)^{\pi(s) - \bar{s}_w} \prod_{j=w+1}^{n-1}p_j$.
Using
\begin{eqnarray}
    &&B_w \frac{1}{\sqrt{2}}(|s\rangle + i^{\theta} |s+e_w+e_{w+1}\rangle)\\
    &=& \frac{1}{2}(|s\rangle - i (-1)^{s_w + s_{w+1}}|s+e_w+e_{w+1}\rangle + i^{\theta}|s+e_w+e_{w+1}\rangle - i^{\theta+1}(-1)^{s_w+s_{w+1}} |s\rangle) \\
    &=& \frac{1}{2}(1-i (-1)^{s_w+s_{w+1}+k})(|s\rangle+i^{\theta}|s+e_w+e_{w+1}) \\
    &=& e^{-i\frac{\pi}{4}(-1)^{s_w+s_{w+1}+k}}\frac{1}{\sqrt{2}}(|s\rangle + i^{\theta} |s+e_w+e_{w+1}\rangle)
\end{eqnarray}
for $\theta = 2k$, $k=0,1$, and
\begin{equation}
    e^{i\frac{\pi}{4}(1-\Gamma)}|s\rangle = |s\rangle,  e^{i\frac{\pi}{4}(1-\Gamma)}|s+e_w+e_{w+1}\rangle = i|s+e_w+e_{w+1}\rangle,
\end{equation}
we have
\begin{eqnarray}
    &&B_w^{b_w} \frac{1}{\sqrt{2}}(|s\rangle + i^{\theta} |s+e_w+e_{w+1}\rangle)\\
    &=& e^{-i\frac{\pi}{4}b_w(-1)^{s_w+s_{w+1}+k}}\frac{1}{\sqrt{2}}(|s\rangle + i^{\theta} |s+e_w+e_{w+1}\rangle)\\
    &=& e^{-i\frac{\pi}{4}b_w(-1)^{s_w+s_{w+1}+k}} e^{-i\frac{\pi}{4}(1-\Gamma)} \frac{1}{\sqrt{2}}(|s\rangle + i^{\theta+1} |s+e_w+e_{w+1}\rangle).
\end{eqnarray}

Since $\Gamma$ commutes with all $B_{j \neq w}$, we can write
\begin{eqnarray}
    &&U_C U_B \frac{1}{\sqrt{2}}(|s\rangle + i^{\theta} |s+e_w+e_{w+1}\rangle) \\
    &=& e^{-i\frac{\pi}{4}b_w(-1)^{s_w+s_{w+1}+k}} U_C e^{-i\frac{\pi}{4}(1-\Gamma)} U_B \frac{1}{\sqrt{2}}(|s\rangle + i^{\theta+1} |s+e_w+e_{w+1}\rangle),
\end{eqnarray}
which after appropriate updates brings us back to the original problem of Step 3 but with odd $\theta \leftarrow \theta +1$ (\textit{Case 1}).
To ensure that we are updating $U_C$ with a control-type gate, we rewrite
\begin{equation}
    e^{-i\frac{\pi}{4}b_w(-1)^{s_w+s_{w+1}+k}} U_C e^{-i\frac{\pi}{4}(1-\Gamma)}
    = e^{-i\frac{\pi}{4} [b_w(-1)^{s_w+s_{w+1}+k} + (1 - (-1)^{\pi(s) - \bar{s}_w})]} U_C e^{-i\frac{\pi}{4}((-1)^{\pi(s) - \bar{s}_w}-\Gamma)},
\end{equation}
knowing that
\begin{equation}
    e^{-i\frac{\pi}{4}((-1)^{\pi(s) - \bar{s}_w}-\Gamma)} |0\rangle = |0\rangle.
\end{equation}
Therefore, the updates are
\begin{eqnarray}
    \phi &\leftarrow& \phi - [b_w(-1)^{s_w+s_{w+1}+k} + (1 - (-1)^{\pi(s) - \bar{s}_w})] \\
    U_C &\leftarrow& U_C e^{-i\frac{\pi}{4}((-1)^{\pi(s) - \bar{s}_w}-\Gamma)},
\end{eqnarray}
while the explicit update rules for the latter (parameters $(\omega, E, F, G)$) can be found in Appendix B.

\subsection{Applying a general Majorana Clifford rotation}

It is easy to generalize the above results to general Majorana Clifford rotations of the form
\begin{equation}
    U = e^{\pm i \frac{\pi}{4} \Gamma}
\end{equation}
where $\Gamma$ is a Hermitian Majorana operator. We follow Step 1 of the previous Section to write
\begin{equation}
    U |\psi\rangle = e^{i\frac{\pi}{4} \phi} U_C U_B \frac{1}{\sqrt{2}} (1 + \Gamma') |s\rangle,
    \label{eq:majorana_rotation_step_1}
\end{equation}
where $\Gamma' = \pm i U_B^{\dagger} U_C^{\dagger} \Gamma U_C U_B$. Now, Eq.~(\ref{eq:majorana_rotation_step_1}) is of the form of Eq.~(\ref{eq:step1}) and the remaining steps are exactly the same. The overall scaling is the same as for $\eta_{jk}$, $\mathcal{O}(n^2)$.

\section{Stabilizer tableau procedures}

\subsection{Matrix identities for $E$, $F$, and $G$}

Here we derive identities (\ref{eq:main_tableau_identity1}) and (\ref{eq:main_tableau_identity2}) satisfied by matrices $E$, $F$, and $G$. We note that $p_i$ and $c_j$ anticommute for $i=j$ and commute otherwise, and this relation remains true after transforming with $U_C$. Therefore, using Eq.~(\ref{eq:commutation}), we can show that
\begin{equation}
    E_{i} \cdot G_{j} = \delta_{ij},
\end{equation}
which can be written in matrix form as
\begin{equation}
    E \cdot G^{T} = I.
\end{equation}
By using the fact that these matrices are full-rank square matrices and thus invertible, we can augment this with additional identities
\begin{equation}
    E \cdot G^{T} = G \cdot E^{T} =E^T \cdot G =G^T \cdot E =I. \label{eq:tableau_identity1}
\end{equation}
Similarly, using Eq.~(\ref{eq:commutation}) and the fact that $c_i$ and $c_j$ anticommute except if $i=j$, we can write
\begin{equation}
    F_i \cdot G_j + G_i \cdot(F_j + G_j) + 1 = \delta_{ij} + 1,
\end{equation}
where the 1 on the left-hand side comes from the fact that both Majorana operators have odd parity and the 1 on the right-hand side ensures that the right-hand side is 1 unless $i=j$. This can again be formulated as a matrix equation:
\begin{equation}
    F \cdot G^T + G \cdot F^{T} = I + G \cdot G^T. \label{eq:tableau_identity2}
\end{equation}
Using (\ref{eq:tableau_identity1}), we also have
\begin{equation}
    E^T \cdot F + F^{T} \cdot E = I + E \cdot E^T.
\end{equation}

\subsection{Update rules for applying $\exp(\pm i \frac{\pi}{4}(I - \Gamma_p))$ from the right}

Let us assume we are given a p-type Majorana operator
\begin{equation}
    \Gamma_p = \prod_{k=0}^{n-1} p_k^{z_k}
\end{equation}
and $U_C$ defined by $(\omega, E, F, G)$. We wish to compute a set of parameters $(\omega', E', F', G')$ that represents
\begin{equation}
    U_C' = U_C e^{\pm i \frac{\pi}{4}(I - \Gamma_p)}.
\end{equation}

Performing the update will not modify $E$ because $[\Gamma_p, p_j]=0$. For the remaining parameters, we compute
\begin{eqnarray}
   e^{i\frac{\pi}{4}(-1)^{\vartheta}(1-\Gamma)} U_C^{\dagger} c_j U_C e^{-i\frac{\pi}{4}(-1)^{\vartheta}(1-\Gamma)}
   &=& e^{-i\frac{\pi}{4}(-1)^{\vartheta}\Gamma} i^{\omega_j} \prod_{k=0}^{n-1} p_k^{F_{jk}} c_k^{G_{jk}} e^{i\frac{\pi}{4}(-1)^{\vartheta}\Gamma}\\
   &=& e^{-i\frac{\pi}{2} (-1)^{\vartheta}(G_j \cdot z) \Gamma} i^{\omega_j}  \prod_{k=0}^{n-1} p_k^{F_{jk}} c_k^{G_{jk}}\\
   &=& \left[ -i (-1)^{\vartheta} \prod_{k=0}^{n-1} p_k^{z_k} \right]^{G_j \cdot z} i^{\omega_j}  \prod_{k=0}^{n-1} p_k^{F_{jk}} c_k^{G_{jk}}\\
   &=& i^{\omega_j - (-1)^{\vartheta} G_j \cdot z } \prod_{k=0}^{n-1} p_k^{F_{jk} + z_k (G_j \cdot z) } c_k^{G_{jk}}.
\end{eqnarray}
The update rules are
\begin{equation}
    \omega_j \leftarrow \omega_j - (-1)^{\vartheta} G_j\cdot z, \quad F_j \leftarrow F_j + z(G_j\cdot z).
\end{equation}

\subsection{Update rules for applying general control gate from the right}

Here, we are given a p-type Majorana operator $\Gamma_1$, a Hermitian, even-parity Majorana operator $\Gamma_2$ that commutes with $\Gamma_1$, and $U_C$ defined by $(\omega, E, F, G)$. We wish to compute $(\omega', E', F', G')$ that represent
\begin{equation}
    U_C' = U_C C_{\Gamma_1, \Gamma_2},
\end{equation}
where
\begin{equation}
    C_{\Gamma_1, \Gamma_2} = e^{-i \frac{\pi}{4} (I - \Gamma_1)(I-\Gamma_2)}
\end{equation}
is a multi-site control gate that multiplies a computational basis state $|t\rangle$ by $\Gamma_2$ if $\Gamma_1|t\rangle = -|t\rangle$.

We begin by deriving the following useful expression
\begin{eqnarray}
    C_{\Gamma_1, \Gamma_2}^{\dagger} \Gamma C_{\Gamma_1, \Gamma_2}
    &=& e^{i\frac{\pi}{4} (I-\Gamma_1) (I-\Gamma_2)} e^{-i\frac{\pi}{4} (I-(-1)^{\alpha_1}\Gamma_1) (I-(-1)^{\alpha_2}\Gamma_2)} \Gamma \\
    &=& e^{i\frac{\pi}{4} [((-1)^{\alpha_1}-1) \Gamma_1 + ((-1)^{\alpha_2}-1] \Gamma_2 + (1-(-1)^{\alpha_1 + \alpha_2})\Gamma_1\Gamma_2]} \Gamma \\
    &=& (-i\Gamma_1)^{\alpha_1} (-i\Gamma_2)^{\alpha_2} (i\Gamma_1\Gamma_2)^{\alpha_1 + \alpha_2} \Gamma \\
    &=& (-1)^{\alpha_1 \alpha_2} \Gamma_1^{\alpha_2} \Gamma_2^{\alpha_1} \Gamma,
    \label{eq:C_transformation}
\end{eqnarray}
where $\alpha_{i}$ is 1 if $[\Gamma_{i},\Gamma]\neq 0$ and 0 otherwise.

Then, to update $E$, we have
\begin{equation}
    U_C^{\prime \dagger} p_j U_C = C_{\Gamma_1, \Gamma_2}^{\dagger} U_C^{\dagger} p_j U_C C_{\Gamma_1, \Gamma_2} = C_{\Gamma_1, \Gamma_2}^{\dagger} \prod_{k=0}^{n-1} p_k^{E_{j,k}} C_{\Gamma_1, \Gamma_2}.
\end{equation}
Let $P_j = \prod_{k=0}^{n-1} p_k^{E_{j,k}}$. We know that $\alpha_{j,1} =0$ because $[\Gamma_1, P_j]=0$ and $\alpha_{j,2} = E_j \cdot x_2$ [using Eq.~(\ref{eq:commutation})]. Therefore, following (\ref{eq:C_transformation}),
\begin{equation}
    C_{\Gamma_1, \Gamma_2}^{\dagger} P_j C_{\Gamma_1, \Gamma_2} = \Gamma_1^{E_j \cdot x_2} P_j = \prod_{k=0}^{n-1} p_k^{E_{j,k} + z_{1,k}(E_j \cdot x_2)}
\end{equation}
and the update rule is
\begin{equation}
    E_j \leftarrow E_j + z_1 (E_j \cdot x_2),\quad j=0,1,\dots, n-1.
\end{equation}

To update $\omega, F, G$, we set
\begin{equation}
    \Gamma^{(j)} = i^{\omega_j} \prod_{k=0}^{n-1} p_k^{F_{jk}} c_k^{G_{jk}}
\end{equation}
and compute $\alpha_{j,1}$ and $\alpha_{j,2}$ using Eq.~(\ref{eq:commutation}). Then, by applying (\ref{eq:C_transformation}) and the formula for the product of Majorana operators [Eq.~(\ref{eq:MajoranaProduct})], we obtain
\begin{equation}
    \Gamma^{(j) \prime} = C_{\Gamma_1, \Gamma_2}^{\dagger} \Gamma^{(j)} C_{\Gamma_1, \Gamma_2}
\end{equation}
whose parameters are the new $(\omega_j, F_j, G_j)$.

\subsection{Stabilizer tableau for $U_C^{\dagger}$}

Let us assume we are given $U_C$ in the form of matrices $E$, $F$, and $G$, and an array $\omega$. Our goal is to find $(\omega', E', F', G')$ such that
\begin{equation}
    U_C p_j U_C^{\dagger} = \prod_{k=0}^{n-1} p_k^{E_{jk}'}, \quad U_C c_j U_C^{\dagger} = i^{\omega_j'} \prod_{k=0}^{n-1} p_k^{F_{jk}'} c_k^{G_{jk}'},
\end{equation}
i.e., the representation of $U_C^{\dagger}$.

To derive the expressions for $(\omega', E', F', G')$, we expand
\begin{eqnarray}
    p_i &=& U_C U_C^{\dagger} p_i U_C U_C^{\dagger} \\
    &=& U_C \prod_{j=0}^{n-1} p_j^{E_{ij}} U_C^{\dagger} \\
    &=& \prod_{j=0}^{n-1} \left( \prod_{k=0}^{n-1} p_k^{E_{jk}'} \right)^{E_{ij}} \\
    &=& \prod_{k=0}^{n-1} p_k^{\sum_{j=0}^{n-1} E_{ij} E_{jk}'},
\end{eqnarray}
which leads to
\begin{equation}
    E \cdot E' = I,
\end{equation}
i.e., $E'$ is matrix inverse of $E$. From Eq.~(\ref{eq:tableau_identity1}) we know that this is equal to $G^T$, i.e.,
\begin{equation}
    E' = G^T.
\end{equation}

We now repeat the same procedure for an arbitrary $c_i$:
\begin{eqnarray}
    c_i &=& U_C U_C^{\dagger} c_i U_C U_C^{\dagger} \\
    &=& U_C i^{\omega_i} \prod_{j=0}^{n-1} p_j^{F_{ij}} c_j^{G_{ij}} U_C^{\dagger} \\
    &=& i^{\omega_i} \prod_{j=0}^{n-1} \left( \prod_{k=0}^{n-1} p_k^{E'_{jk}} \right)^{F_{ij}} \left(  i^{\omega_j'}\prod_{k=0}^{n-1} p_k^{F_{jk}'} c_k^{G_{jk}'}\right)^{G_{ij}} \\
    &=& i^{\omega} \prod_{k=0}^{n-1} p_k^{\sum_{j=0}^{n-1}F_{ij} E'_{jk} + G_{ij} F_{jk}'} c_k^{\sum_{j=0}^{n-1}G_{ij} G_{jk}'},
\end{eqnarray}
where in the last line we chose to store all phases and sign changes due to anticommutation into a new variable $\omega$. We can now read off
\begin{equation}
    G \cdot G' = I, \quad F\cdot E' + G \cdot F' = 0.
\end{equation}
From this, and using identities derived earlier, we can obtain
\begin{eqnarray}
    G' &=& G^{-1} = E^{T}, \\
    F' &=& G^{-1} \cdot F \cdot E' = E^T \cdot F \cdot G^{T}.
\end{eqnarray}
Finally, to recover $\omega'$, we note that
\begin{equation}
     c_i = U_C^{\dagger} U_C c_i U_C^{\dagger} U_C = i^{\omega_i'} \Gamma_i,
\end{equation}
where, by construction, we know that
\begin{equation}
    \Gamma_i \propto c_i.
\end{equation}
We can therefore initially set $\omega_i'=0$ and $(E', F', G')$ as derived above, compute $\Gamma_i$ using this definition of $U_C^{\dagger}$, and then extract $\omega_i'$ as the negative of the phase factor of the evaluated $\Gamma_i$.

The overall cost of computing the representation of $U_C^{\dagger}$ is $\mathcal{O}(n^3)$ due to matrix multiplications and final evaluation of $\omega'$ array.

\subsection{Action on a computational basis state}

We now wish to evaluate $\theta$ and $t$ in
\begin{equation}
    U_C^{\dagger} |s\rangle = i^{\theta} |t\rangle,
\end{equation}
where $U_C$ is defined by its stabilizer tableux and $s$ is a bitstring defining a computational basis state. We use
\begin{eqnarray}
    U_C^{\dagger} |s\rangle = U_C^{\dagger} \prod_{j=0}^{n-1} c_i^{s_i} |0\rangle = U_C^{\dagger} \prod_{j=0}^{n-1} c_i^{s_i} U_C U_C^{\dagger} |0\rangle = \Gamma |0\rangle,
\end{eqnarray}
where $\Gamma =U_C^{\dagger} \prod_{j=0}^{n-1} c_i^{s_i} U_C$ can be evaluated using the stabilizer tableau in $\mathcal{O}(|s|n) = \mathcal{O}(n^2)$ time and $\Gamma |0\rangle$ is given in Eq.~(\ref{eq:gamma_times_s}).

\end{document}